\documentclass[aps,prc,preprint,groupedaddress]{revtex4}

\newcommand{\beq}{\begin{equation}}
\newcommand{\eeq}{\end{equation}}
\newcommand{\beqa}{\begin{eqnarray}}
\newcommand{\eeqa}{\end{eqnarray}}

\usepackage{epsfig}
\usepackage{youngtab}

\begin{document}

 \title{The role of $q\bar q$ components in the
$N(1440)$ resonance}

\author{Q. B. Li}
\email[]{ligb@pcu.helsinki.fi}
\affiliation{Helsinki Institute of Physics
POB 64, 00014 University of Helsinki, Finland}

\author{D. O. Riska}
\email[]{riska@pcu.helsinki.fi}
\affiliation{Helsinki Institute of Physics, POB 64,
00014 University of Helsinki, Finland}

\thispagestyle{empty}

\date{\today}
\begin{abstract}
The role of
5-quark components in the 
pion and electromagnetic decays and transition form
factors of the $N(1440)$ resonance is explored.
The $qqqq\bar q$ components, where the 4-quark
subsystem has the flavor-spin symmetries
$[4]_{FS}[22]_F[22]_S$ and $[4]_{FS}[31]_F[31]_S$, which are
expected to have
the lowest energy of all $qqqq\bar q$ configurations,
are considered in detail with
a nonrelativistic quark model.
The matrix elements 
between the 5-quark components of the $N(1440)$ and
the nucleon,
$qqqq\bar q\rightarrow qqqq\bar q$, play a minor role in  
these decays, while the transition 
matrix elements $qqqq\bar q \rightarrow qqq$ 
and $qqq\rightarrow qqqq\bar q$ that involve
quark antiquark annihilation are very significant.
Both for electromagnetic and strong decay
the change from the
valence quark model value is dominated by the 
confinement triggered $q\bar q$ annihilation transitions.
In the case of pion decay the calculated decay width
is enhanced substantially both by the direct 
$q\bar q\rightarrow \pi$ and also by 
the confinement triggered $q\bar q\rightarrow \pi$ transitions.
Agreement with the empirical value for the
pion decay width may be reached with a $\sim 30$\% $qqqq\bar q$ 
component in the $N(1440)$.

\end{abstract}

\pacs{}

\maketitle

\section{Introduction}

The conventional 3-quark model for nucleons and nucleon
resonances leads to values for both the electromagnetic
and the strong decay widths of the lowest nucleon resonance,
$(N(1440))$, which are smaller by an order of magnitude
than the corresponding experimental values
\cite{bruno, melde}. This suggests that this resonance
should have significant $q\bar q$, if not more
exotic components, in addition
to the 3 valence quarks. This is also suggested by the 
fact that this resonance appears naturally at its very low
energy as a vibrational - i.e collective - state in the
Skyrme model \cite{bied}.
This issue is studied here by explicit consideration
of the $qqqq\bar q$ configurations, which are expected
to have the lowest energy, and which therefore may be
expected to form a significant components in this
resonance. 

Admixture of a $qqqq\bar q$ component with a
probability of $\sim$ 30\% is found to increase the
calculated helicity amplitude $A_{1\over 2}$ for 
$N(1440)\rightarrow N\gamma$ decay moderately and the width
of $N(1440)\rightarrow N\pi$ decay significantly. 
While the matrix elements 
between the 5-quark components of the $N(1440)$ and
the nucleon,
$qqqq\bar q\rightarrow qqqq\bar q$, play but a minor role in  
these decays, the transition 
matrix elements $qqqq\bar q \rightarrow qqq$ 
and $qqq\rightarrow qqqq\bar q$ that involve
quark antiquark annihilation are very significant. 
The calculated helicity amplitude and the modification
from the
valence quark model value is very sensitive to the 
transitions between the $qqqq\bar q$ and $qqq$ 
components that are triggered 
by the confinement between quarks.
In the case of pion decay the calculated decay width
is substantially enhanced by the direct 
transition and also by 
the confinement triggered transitions 
between the $qqqq\bar q$ and $qqq$ 
components. It is found that with a $\sim$30\% $qqqq\bar q$ component 
in the $N(1440)$ it becomes possible to reach agreement with the 
empirical pion decay width. It is also found that 
recent data on the
shape of the empirical helicity amplitude $A_{1\over 2}$  for
the $N(1440)\rightarrow N \gamma$ transition demand a 
spatially extended wave function model. 

This paper is arranged as follows:
The $qqqq\bar q$ wave functions of the proton and
the $N(1440)$ are described in Section II.
In section III the electromagnetic decay of the $N(1440)$ resonance 
is studied by explicit calculation of
the effect of transitions between the $qqqq\bar q$ components 
of the proton and the $N(1440)^+$ and the corresponding 
annihilation transitions
$qqqq\bar q\rightarrow qqq\gamma$ and $qqq\rightarrow qqqq\bar q\gamma$.
In Section IV the strong decay of the $N(1440)^+$ is considered 
along the lines of the electromagnetic decay.
Section IV contains a concluding
discussion. 

\section{The wave functions of nucleon and $N(1440)$}

With addition of a $qqqq\bar q$ component in addition to the 
conventional $qqq$ configuration, the complete wave functions 
of the nucleon and the $N(1440)$ resonance may be written
schematically in the form:
\beqa
&&\vert N,s_z \rangle=A_{p3}\vert N, s_z\rangle_{3q}+
A_{p5}\vert N,s_z \rangle_{5q}\, , \nonumber\\
&&\vert N(1440),s_z \rangle=A_{r3}\vert N(1440),s_z \rangle_{3q}+
A_{r5}\vert N(1440),s_z \rangle_{5q} .
\label{nrwave}
\eeqa
Here $A_{p3}$ and $A_{p5}$ are are the amplitude
factors for the $qqq$ and 
$qqqq\bar q$ components in the nucleon and
$A_{r3}$ and $A_{r5}$ 
are the corresponding factors for the
$N(1440)$, respectively. Their spin projection on the
$z-$axis is denoted $s_z$. 
The matrix elements of the amplitudes for electromagnetic and strong
decay will be formed of diagonal
transitions terms: $qqq\rightarrow qqq$, 
$qqqq\bar q\rightarrow qqqq\bar q$ as well as of 
non-diagonal terms of the type $qqqq\bar q\rightarrow qqq$
and $qqq\rightarrow qqqq\bar q$,
which involve quark-antiquark annihilation.

\subsection{The wave functions of the $qqq$ components}
In the harmonic oscillator quark model the wave functions of 
the nucleon and the $N(1440)$ in 
the $qqq$ configuration have the 
standard expressions:
\beqa
&&\vert N,s_z \rangle_{3q}={1\over\sqrt{2}}~\bigg(
\vert {1\over 2},t_z\rangle_+
\vert {1\over 2},s_z\rangle_+ +
\vert {1\over 2},t_z\rangle_-
\vert {1\over 2},s_z\rangle_-\bigg)
\phi_{000}(\vec\xi_2)\phi_{000}(\vec \xi_1),\nonumber\\
&&\vert N(1440),s_z \rangle_{3q}={1\over\sqrt{2}}~\bigg(
\vert {1\over 2},t_z\rangle_+
\vert {1\over 2},s_z\rangle_+ +
\vert {1\over 2},t_z\rangle_-
\vert {1\over 2},s_z\rangle_-\bigg) \nonumber\\
&&~~~~~~~~~~~~~~~~~~~~~~
\cdot {1\over\sqrt{2}}~\bigg(\phi_{200}(\vec\xi_2)\phi_{000}(\vec \xi_1)
+\phi_{000}(\vec\xi_2)\phi_{200}(\vec \xi_1)\bigg).
\label{3qwave}
\eeqa
Here $\vert {1\over 2},s_z\rangle_\pm$ and 
$\vert {1\over 2},t_z\rangle_\pm$, 
with $t_z$ being the isospin-z component, are spin and 
isospin wave functions 
of mixed symmetry ($[21]$), in which (+) denotes a state
that is symmetric ``(112)''  
and (-) denotes a 
state that is antisymmetric ``(121)'' under exchange of the
spin or isospin of the 
first two quarks.
The Jacobi coordinates 
$\vec \xi_1$ and $\vec \xi_2$
are defined by the 3-quark position coordinates 
as:
\beqa
\vec \xi_1 ={1\over \sqrt{2}}({\vec r}_1 - {\vec r}_2)~,~~~~~
\vec\xi_2 ={1\over \sqrt{6}}({\vec r}_1 + {\vec r}_2 - 2{\vec r}_3)\,.
\eeqa
The harmonic oscillator wave functions are:  
\beqa
\phi_{000}(\vec\xi_i)=({\omega_3^2\over \pi})^{{3\over 4}}
~ e^{-\omega_3^2 \xi_i^2/2},~~
\phi_{200}(\vec\xi_i)=\sqrt{{2\over 3}}\omega_3^2(\xi_i^2-
{3\over {2\omega_3^2}})\phi_{000}(\vec\xi_i)\, .
\label{harm3q}
\eeqa
Here the subscripts denote the quantum numbers $(nlm)$ 
of the oscillator wave functions. 

\subsection{The wave functions of the $qqqq\bar q$ components}

Positive parity demands that 
the $qqqq\bar q$ admixtures in the proton and the
$N(1440)^+$ 
be $P-$wave states (or states with higher odd angular
momentum). By the conventional assumption that
the hyperfine interaction between the
quarks is spin-dependent, it follows that the $qqqq$
configurations that have the lowest energy is that,
for which the spin state has the highest possible
degree of antisymmetry \cite{helminen}.
A similar argument applies in the case the hyperfine
interaction is flavor dependent.
The simplest hyperfine interaction model, which leads
to a realistic splitting between the nucleon and the
$N(1440)$ is 
the schematic flavor and spin dependent hyperfine 
interaction between the
quarks $-C_\chi\sum_{i<j} \vec\lambda_F^i\cdot
\vec\lambda_F^j\,\vec\sigma_i\cdot\vec\sigma_j$, where $C_\chi$
is a constant parameter ($C_\chi\sim 20-30$ MeV)
\cite{glozman}. This implies that the $qqqq\bar q$ configuration that
has the lowest energy, and which is most likely to
form notable admixtures in the nucleons and the
$N(1440)$, has the mixed spin-flavor symmetry 
$[4]_{FS}[22]_F[22]_S$, with the antiquark in the ground state
\cite{helminen}. The flavor-spin symmetry of the
state with the next to lowest energy is
$[4]_{FS}[31]_F[31]_S$.

The $qqqq\bar q$ component in the proton with
spin-flavor symmetry $[4]_{FS}[22]_F[22]_S$   
has mixed spatial symmetry $[31]_X$, and may be
represented by the wave function:

\begin{eqnarray}
&&\vert N,s_z \rangle_{5q}={1\over\sqrt{2}}\sum_{a,b}
\sum_{m,s} (1,1/2,m,s\vert\, 1/2,s_z)) C^{[1^{4}]}_{[31]_{a}
[211]_{a}}\,
\nonumber\\
&&[211]_C(a)\,[31]_{X,m}(a)\, [22]_F(b)\,[22]_{S}(b)\, \bar\chi_{t_z,s}\,
\psi(\{{\vec \xi}_i\})\, .
\label{5qn}
\end{eqnarray}
Here the color, space and flavor-spin wave functions of the
$qqqq$ subsystem have been denoted by their Young patterns
respectively, and the sum over $a$ runs over the 3 
configurations of the $[211]_C$ and $[31]_X$ representations
of $S_4$, and the sum over $b$ runs over the 2 
configurations of the
$[22]$ representation of $S_4$ respectively \cite{chen}.
Here $C^{[1^{4}]}_{[31]_{a}
[211]_{a}}$ denotes the $S_4$ Clebsch-Gordan coefficients
of the representations $[1111][31][211]$.
  
The explicit color-space part of the
wave function (\ref{5qn})
may be expressed in the form:
\beqa
&&\psi_C(\{{\vec\xi}_i\})={1\over \sqrt{3}}
\{C_1~\varphi_{01m}(\vec\xi_1)\varphi_{000}(\vec\xi_2)
\varphi_{000}(\vec\xi_3)\varphi_{000}(\vec\xi_4)-\nonumber\\
&&C_2~\varphi_{000}(\vec\xi_1)\varphi_{01m}(\vec\xi_2)
\varphi_{000}(\vec\xi_3)\varphi_{000}(\vec\xi_4) 
 +~C_3 ~\varphi_{000}(\vec\xi_1)\varphi_{000}(\vec\xi_2
)\varphi_{01m}(\vec\xi_3)\varphi_{000}(\vec\xi_4)~\}.
\label{5qno}
\eeqa
Here the additional coordinate vectors $\vec\xi_i$, i=3,4, are 
defined as: 
\begin{eqnarray}
&&\vec \xi_3 ={1\over\sqrt{12}}(\vec r_1 +\vec r_2 +\vec r_3
 - 3\vec r_4)\, ,\nonumber\\
&&\vec\xi_4 = {1\over\sqrt{20}}(\vec r_1 +\vec r_2 + \vec r_3
+\vec r_4 - 4 \vec r_5)\, ,
\label{relative}
\end{eqnarray}
and the oscillator wave functions are defined as:
\beqa
\varphi_{000}({\vec\xi}_i)=({\omega_5^2\over \pi})^{{3\over 4}}
~ e^{-\omega_5^2 \xi_i^2/2},~~
\varphi_{01m}({\vec\xi}_i)=\sqrt{2}\omega_5 \xi_{i,m}\, 
\varphi_{000}({\vec\xi_i}).
\label{harm5q0}
\eeqa
Note that the oscillator parameter for the $qqqq\bar q$
component $\omega_5$ may differ from that in  the
$qqq$ component (\ref{harm3q}). In Eq. (\ref{5qno})
$C_i$ (i=1, 2, 3) represent the color wave functions of
the 3 configurations of $[211]_C$ and notice has been taken 
that the vectors $\vec\xi_i$ (i=1, 2, 3) realize the 3 
symmetry configurations
of $[31]_X$ in orbital space \cite{chen}.   

In the present model, the wave function of the 5-quark 
component of $N(1440)$ has the same structure as 
the nucleon in the 
spin-flavor-color space. Orthogonality against the nucleon 
determines the corresponding wave function in the orbital space. 
Schematically 
the $N(1440)$ wave function in the $qqqq\bar q$ configuration with 
the lowest energy may be written as \cite{helminen}:
\begin{eqnarray}
&&\vert N(1440),s_z \rangle_{5q}={1\over\sqrt{2}}\sum_{a,b}
\sum_{m,s} (1,1/2,m,s\vert\, 1/2,s_z)C_{[31]_a [211]_a}
^ {[1^ 4]}\,
\nonumber\\
&&[211]_C(a)\,[31]_{X,m}(a)\, [22]_F(b)\,[22]_{S}(b)\, \bar\chi_{t_z,s}\,
\Psi(\{\vec\xi_i\})\, .
\label{5qr}
\end{eqnarray}
Here the orbital wave function of the $qqqq\bar q$ component 
$\Psi(\{\vec\xi_i\})$ 
lies in the $n=2$ band and is combined with the color wave 
functions $[211]_C$
and the spatial wave functions $[31]_X$ in the following way:
\beqa
&&\Psi_C(\{{\vec\xi}_i\})={1\over 2\sqrt{3}}
\{C_1~[\varphi_{21m}(\vec\xi_1)\varphi_{000}(\vec\xi_2)\varphi_{000}
(\vec\xi_3)\varphi_{000}(\vec\xi_4)
+\varphi_{01m}(\vec\xi_1)\varphi_{200}(\vec\xi_2)\varphi_{000}
(\vec\xi_3)\varphi_{000}(\vec\xi_4)\nonumber\\
&&+\varphi_{01m}(\vec\xi_1)\varphi_{000}(\vec\xi_2)\varphi_{200}
(\vec\xi_3)\varphi_{000}(\vec\xi_4)
+\varphi_{01m}(\vec\xi_1)\varphi_{000}(\vec\xi_2)\varphi_{000}
(\vec\xi_3)\varphi_{200}(\vec\xi_4)] ~-~\nonumber\\
&&C_2~[\varphi_{200}(\vec\xi_1)\varphi_{01m}(\vec\xi_2)\varphi_{000}
(\vec\xi_3)\varphi_{000}(\vec\xi_4)
+\varphi_{000}(\vec\xi_1)\varphi_{21m}(\vec\xi_2)\varphi_{000}
(\vec\xi_3)\varphi_{000}(\vec\xi_4)\nonumber\\
&&+\varphi_{000}(\vec\xi_1)\varphi_{01m}(\vec\xi_2)\varphi_{200}(\vec\xi_3)
\varphi_{000}(\vec\xi_4)
+\varphi_{000}(\vec\xi_1)\varphi_{01m}(\vec\xi_2)\varphi_{000}
(\vec\xi_3)\varphi_{200}(\vec\xi_4)] ~+~\nonumber\\
&&C_3~[\varphi_{200}(\vec\xi_1)\varphi_{000}(\vec\xi_2)\varphi_{01m}
(\vec\xi_3)\varphi_{000}(\vec\xi_4)
+\varphi_{000}(\vec\xi_1)\varphi_{200}(\vec\xi_2)\varphi_{01m}
(\vec\xi_3)\varphi_{000}(\vec\xi_4)\nonumber\\
&&+\varphi_{000}(\vec\xi_1)\varphi_{000}(\vec\xi_2)\varphi_{21m}
(\vec\xi_3)\varphi_{000}(\vec\xi_4)
+\varphi_{000}(\vec\xi_1)\varphi_{000}(\vec\xi_2)\varphi_{01m}
(\vec\xi_3)\varphi_{200}(\vec\xi_4)]\}.
\label{5qro}
\eeqa 
The additional harmonic oscillator wave functions
here are defined as:
\beq
\varphi_{200}({\vec\xi}_i)=\sqrt{2\over 3}\omega_5^2
(\xi_i^2-{3\over 2\omega_5^2})\varphi_{000}({\vec\xi_i}),~~
\varphi_{21m}({\vec\xi}_i)={2\over \sqrt{7}}\omega_5^3\,\xi_{i,m}
(\xi_i^2-{3\over 2\omega_5^2})\varphi_{000}({\vec\xi_i}).
\label{harm5q1}
\eeq 

In the case of the flavor-spin symmetry configuration
$[4]_{FS}[31]_F[31]_S$ the wave function of the
$qqqq\bar q$ component of the nucleon takes the form:
\beqa
&&\Psi_{[31]}^{(J)}(s_z) ={1\over \sqrt{3}}\sum_{a,b}
\sum_{m,s,M,j;T,t}
(1,1,m,M\vert\, J,j)(J,1/2,j,s\vert\, 1/2,s_z)\,
C^{[1^4]}_{[211]a,[31]a}\,
\nonumber\\
&&(1,1/2,T,t|3/2,1/2)[211]_C(a)\,[31]_{X,m}(a)\, [31]_{F,T}(b)
\,[31]_{S,M}(b)\, \bar\chi_{t,s}\psi(\{\vec\xi_i\}) \,.
\label{q531}
\eeqa
Here $J$ denotes the total angular momentum of the
$qqqq$ system, which takes the values 0 and 1. 
The sum over $a$ again runs over the 3 
configurations of the $[211]_C$ and $[31]_X$ representations
of $S_4$. The sum over $b$ runs over the 3 
configurations of the $[31]$ representation.  
Here the isospin-z component of the 4-quark state is
denoted $T$ and that of the antiquark $t$. 
The corresponding wave function for the 
excited $qqqq\bar q$
component is obtained by replacement of the
spatial wave function $\psi(\{\vec\xi_i\})$ by
the corresponding wave function
$\Psi(\{\vec\xi_i\})$ (\ref{5qr}).

\section{The electromagnetic decay $N(1440)^+\rightarrow p\gamma$}
\subsection{$qqq\rightarrow qqq\gamma$ matrix elements }

Consider $N(1440)^+\rightarrow p\gamma$ decay that is driven
by the direct electromagnetic coupling to individual 
constituent quarks ($qq\gamma$). In the non-relativistic
approximation the matrix elements of the elastic 
and annihilation matrix elements of the current operator
are then:
\begin{eqnarray}
&&\langle \vec p\,'\,\vert\, \vec j \,\vert\, \vec p \,\rangle_{elas}
= {\vec p\,' + \vec p\over 2 m}
+{i\over 2 m}(\vec \sigma \times \vec q)\, ,
\nonumber\\
&&\langle \vec p\,'\,\vert\, \vec j\, \vert\, \vec p\, \rangle_{anni}
= \vec \sigma \, , 
\label{current} 
\end{eqnarray} 
respectively. For pointlike quarks
the electromagnetic transition
operator for elastic transitions between
states with $n_q$ quarks is then: 
\beqa
T=\sum_{i=1}^{n_q}~\sqrt{2}~\frac{e_i}{2m}{{\sigma_i}_-} k_\gamma\,.
\label{dt}
\eeqa
Here $e_i$ and $m$ are the electric charge and mass of the 
quark that emits the photon, respectively, and $n_q$ is the 
number of constituent quarks. Here the momentum
of the final right-handed virtual-photon is taken to be 
in the direction of the
$z$-axis: $\vec k =(0,0,k_\gamma)$ in the center of 
mass frame of the $N(1440)$. It is related to the magnitude 
of four-momentum transfer $Q=\sqrt{-k^2}$ as:
\beqa
k_\gamma^2=Q^2+{({M_R^2-M_N^2-Q^2)^ 2}\over{4M_R^2}}
\label{kgamma},
\eeqa
where $M_R$ and $M_N$ are masses of the $N(1440)$ and the 
nucleon, respectively.
In Eq. (\ref{dt}) the term that is proportional
to quark momentum has been omitted as it does not contribute to 
the transition under consideration. 

The $N(1440)^+\rightarrow p\gamma$ transition is described 
by the helicity amplitude,
\beqa
A_{\frac{1}{2}}={1\over{\sqrt{2 K_\gamma}}}
\langle p,~\frac{1}{2},~ -\frac{1}{2}~| 
~T~ |~N(1440)^+,~\frac{1}{2},~\frac{1}{2}\rangle~,
\label{helam}
\eeqa
for the helicity component $1/2$ of the $N(1440)$ 
on the direction of the photon momentum. Here $K_\gamma$ = 414 MeV 
is the real-photon three momentum in the center of mass 
frame of $N(1440)$.  
 
\begin{figure}[t]
\vspace{20pt} 
\begin{center}
\mbox{\epsfig{file=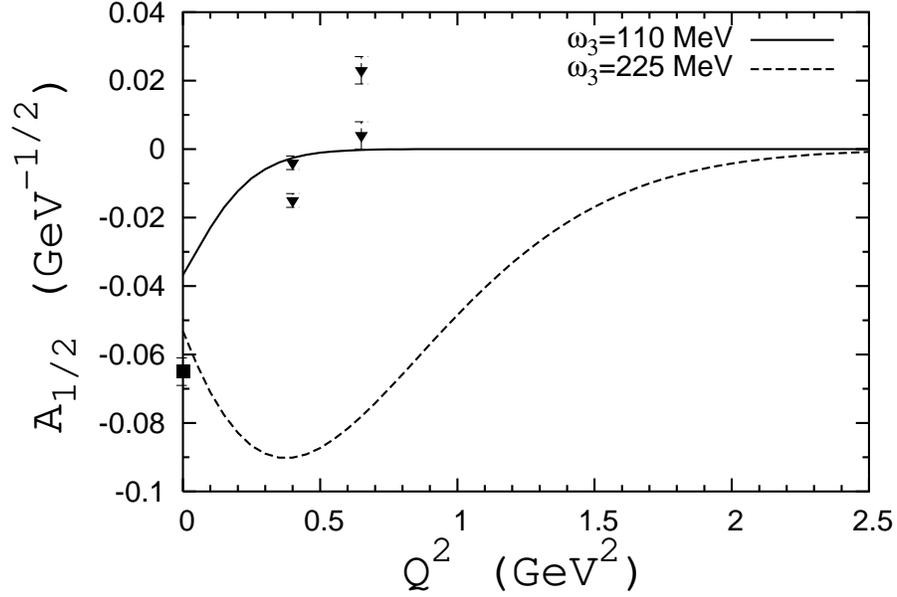, width=120mm}} 
\caption{The helicity amplitude for  
$N(1440)^+\rightarrow p\gamma$ 
in the $qqq$ model for 2 values of the oscillator
frequency $\omega_3$.
The data point at $Q^2 = 0$ is
from \cite{PDG} (square) and the other points are taken from the
phenomenological analysis in \cite{Azna} (triangles).}
\label{amp3q}
\end{center}
\vspace{10pt} 
\end{figure}

The helicity amplitude for $N(1440)^+\rightarrow p\gamma$ decay in 
the conventional $qqq$ configuration may be calculated 
from Eq. (\ref{3qwave}) as:
\beqa
A_{\frac{1}{2}}^{(3q)}=-\frac{\sqrt{3}}{18}~
{k_\gamma^2\over \omega_3^2}~
e^{-{k_\gamma^2/ 6\omega_3^2}}~
\frac{e}{2m} {k_\gamma\over\sqrt{K_\gamma}}.
\label{hel3q}
\eeqa
It is instructive to consider this expression
as a function of the parameter values. It follows
from the expression that the main sensitivity
of the calculated helicity amplitude is to the
oscillator frequency, which is inversely proportional
to the spatial extent of the wave function.
This is illustrated in Fig. \ref{amp3q}, where the
helicity amplitude (\ref{hel3q}) is shown as a function
of $Q^2$ for two values of $\omega_3$
when the constituent quark mass is taken to be $m$ = 340 MeV.
The parameter values 
are $\omega_3=110$ MeV and $\omega_3=225$ MeV,
the latter set by the empirical radius of the
proton as $\omega_3 = 1/r_p$. The data points in the
figure are those given in ref. \cite{Azna}.
It is clear that the empirical shape of the helicity
amplitude is much better recovered with the smaller
value 110 MeV of $\omega_3$. This suggests that
the $N(1440)$ has a much more extended structure than
the nucleon. We shall employ this smaller value
for $\omega_3$ below.
At the photon
point, $Q^2=0$, $k_\gamma=K_\gamma$ and
the expression Eq. (\ref{hel3q}) leads to 
the value $A_{\frac{1}{2}}^{(3q)}=
-0.037/\sqrt{\mathrm{GeV}}$, which is smaller than the corresponding
experimental value $ A_{\frac{1}{2}} =
-0.065\pm0.004/\sqrt{\mathrm{GeV}}$ by a factor 1.8 
\cite{PDG}. 

\subsection{$qqqq\bar q\rightarrow qqqq\bar q\gamma$ transition
matrix elements}

The $qqqq\bar q\rightarrow qqqq\bar q\gamma$ transition 
matrix elements are the matrix elements of the
transition operator (\ref{dt})
between the wave functions (\ref{5qn}) and 
(\ref{5qr}) that represent the
$qqqq\bar q$ components of the nucleon
and the $N(1440)$ resonance respectively. 
The mixed symmetry configuration $[22]_S$ 
of the $qqqq$ subsystem has total spin 0, and as a consequence
the matrix elements of the
current operators of these quarks vanish and do
not contribute to the direct transition 
$N(1440)\rightarrow N$.
Therefore the $qqqq\bar q\rightarrow qqqq\bar q\gamma$ transition 
amplitude only comes from the direct transitions between 
the antiquarks: 
$\bar q\rightarrow \bar q\gamma$, the transition operator of which 
has the opposite sign to that of $q\rightarrow q\gamma$. 

For these,  
the matrix element of the spin-flavor-color part of the
operator is found to be
$C_{SFC}^{(5q)}=1/9$. The matrix element in orbital space calculated 
with the wave functions given by Eqs. (\ref{5qno}) 
and (\ref{5qro}) becomes: 
\beq
\langle\psi_C(\{\xi_i\})\vert e^{-i\vec k_\gamma\cdot {\vec r}_5}
\vert \Psi_C(\{\xi_i\})\rangle=-{\sqrt{6}\over 30}
{k_\gamma^2\over \omega_5^2}~
e^{-{k_\gamma^2/ 5\omega_5^2}}~\, .
\label{5qmeo}
\eeq   
The contribution from  these diagonal $qqqq\bar q$ matrix elements to
the helicity amplitude for $N(1440)^+\rightarrow p\gamma$ decay is 
then found to be:
\beqa
A_{\frac{1}{2}}^{(5q)}=-\frac{\sqrt{6}}{270}~
{k_\gamma^2\over \omega_5^2}~
e^{-{k_\gamma^2/ 5\omega_5^2}}~
\frac{e}{2m} {k_\gamma\over\sqrt{K_\gamma}}.
\label{hel5q}
\eeqa
If the value of the oscillator parameter of the
$qqqq\bar q$ component is taken to be $\omega_5$ = 200 MeV
the numerical value
for this contribution
$A_{\frac{1}{2}}^{(5q)}$ to the helicity
amplitude is very small:
$-0.0047/\sqrt{\mathrm{GeV}}$ at the photon point
$Q^2=0$. This value is much smaller than that 
given by the conventional $qqq$ model. The reason for
this small value is due to the fact that for the
configuration with spin symmetry $[22]$ the 4 quarks
do not contribute to the decay amplitude, so that
the whole contribution arises from the single
antiquark. The result is not very sensitive to the
value of $\omega_5$. 

With the normalizations of the wave functions of nucleon 
and $N(1440)$ in (\ref{nrwave}), the diagonal
contributions to the helicity 
amplitude for $N(1440)^+\rightarrow p\gamma$ is 
obtained by addition of the 
$qqqq\bar q\rightarrow qqqq\bar q\gamma$ and 
the conventional $qqq\rightarrow qqq\gamma$ amplitudes:
\beqa
A_{\frac{1}{2}}=A_{p3}A_{r3}A_{\frac{1}{2}}^{(3q)}
+A_{p5}A_{r5}A_{\frac{1}{2}}^{(5q)}.
\eeqa
As the 5-quark helicity amplitude in (\ref{hel5q}) is much 
smaller than the conventional 3-quark helicity 
amplitude in (\ref{hel3q}), the combination of these
two mechanisms result in a final helicity
amplitude, which is in worse disagreement with 
the empirical value than the value given
by the $qqq$ model, even 
with assumption of a large component of $qqqq\bar q$ in 
the $N(1440)$ resonance.
For instance, assuming a 10\% proportion of the 5-quark component
in proton and 30\% in $N(1440)$, or equivalently
the normalizations
$A_{p3}=0.95$ and $A_{r3}=0.84$, leads to 
$A_{\frac{1}{2}}=-0.030/\sqrt{\mathrm{GeV}}$ at $Q^2=0$ if the phase 
of the normalizations $A_{p5}$ and $A_{r5}$ are assumed to be $+1$, 
which is smaller
than the empirical value by a factor $\sim$ 2.

\subsection{Transitions through $q\bar q$ annihilation}

\subsubsection{Direct $qqqq\bar q\rightarrow qqq\gamma$ and 
$qqq\rightarrow qqqq\bar q\gamma$ transitions}

\begin{figure}[t]
\vspace{20pt} 
\begin{center}
\mbox{\epsfig{file=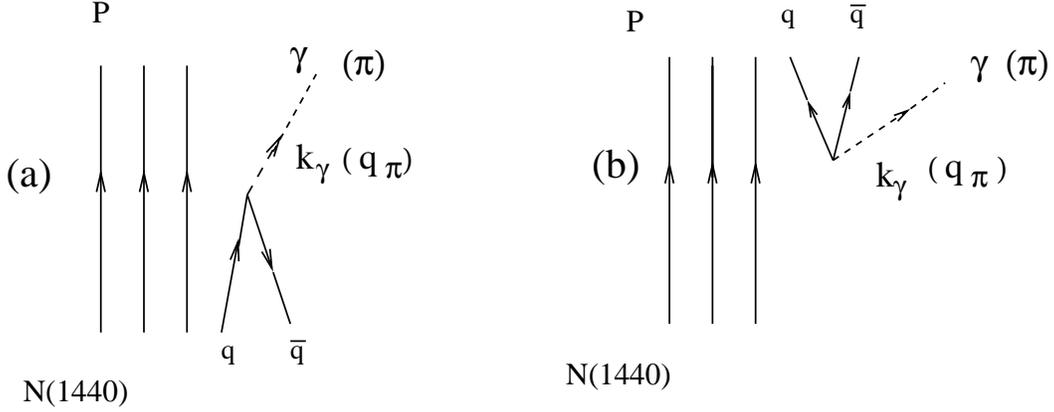, width=140mm}} 
\caption{Direct $q\bar q\rightarrow \gamma (\pi)$ annihilation
process.}
\label{fig1}
\end{center}
\vspace{10pt} 
\end{figure}

In the case of $q\bar q\rightarrow\gamma$
transitions the $\gamma_\mu$ coupling for pointlike quarks
(cf. (\ref{current})) leads to 
the following $q\bar q\rightarrow \gamma$
transition operator that involves annihilation
of $q\bar q$ pair into a right handed photon (Fig. \ref{fig1}):
\beqa
T_{a}=\sum_{i=1}^4 \sqrt{2}~e_i {\sigma_i}_-\, .
\label{at}
\eeqa
Here $e_i$ is the electric charge of the annihilating
quark and ${\sigma_i}_-$ is the spin lowering operator.

For the case of the $qqqq\bar q\rightarrow qqq\gamma$ transition
in Fig.\ref{fig1}(a),
the calculation of the matrix element of 
the operator (\ref{at})
involves calculation of the overlap of the
$qqq$ component of the proton with the residual
$qqq$ component that is left in the $N(1440)^+$
after the annihilation of a $d\bar d$ pair into a photon. It
also requires specification of the spatial
part of the $N(1440)^+$ wave function.
The spin-flavor-color 
matrix element that is derived from the 
wave functions of the proton in (\ref{3qwave}) and 
$N(1440)^+$ in (\ref{5qr}) is:
\beq
C_{SFC}^{(5q\rightarrow 3q)}=-{2\sqrt{3}\over 9}.
\label{sfan}
\eeq 

The following step is the calculation
of orbital matrix element of the direct annihilation 
transition amplitude. This may be cast in the form:
\beq
C_{O}^{(5q\rightarrow 3q)}(k_\gamma)=\langle ~
\phi_{000}({\vec\xi}_1)\phi_{000}({\vec\xi}_2)[111]_C~\vert~
\delta({\vec r}_4 - {\vec r}_5)~
e^{-i{\vec k}_\gamma\cdot {{\vec r}_4 + {\vec r}_5\over 2}}~\vert~
\Psi_C(\{{\vec\xi_i}\})~\rangle,
\label{oan}
\eeq 
where $[111]_C$ denotes the color singlet state of 
the final proton in
in the $qqq$ configuration. Note that only the color 
symmetry configuration of the $qqqq\bar q$ component of 
initial $N(1440)^+$
that is described by $C_3$ or, 
\beq
{\young(14,2,3)}\, , 
\eeq 
in Eq. (\ref{5qro}) leads to a nonzero matrix element, 
when multiplied with the color singlet of the 
$qqq$ component of the proton upon annihilation
of the 4-th quark with the 5-th antiquark in $N(1440)$. 
Hence only the term that is multiplied by $C_3$ 
in Eq. (\ref{5qro}) 
contributes to the matrix element ($\ref{oan}$).
Consequently the matrix element reduces to:
\beq
C_{O}^{(5q\rightarrow 3q)}(k_\gamma)=({\omega_3\omega_5\over\omega^2})^3
\{{\sqrt{3}\over 4}[({\omega_5\over\omega})^2-1]+
({25\over 64}X_1-{3\over 16}X_2-
{3\over 128}X_1 {k_\gamma^2\over \omega_5^2})\}
{k_\gamma\over\omega_5}~e^{-{3\over 20}{k_\gamma^2\over\omega_5^2}}\,.
\label{orban}
\eeq
Here the $X_1$ and $X_2$ are defined as the 
constants $X_1={2\over\sqrt{7}}+{6\over 5\sqrt{3}}$ and 
$X_2={2\over\sqrt{7}}+{2\over \sqrt{3}}$.
In the derivation of (\ref{orban}) the normalization
factor mismatch between the $qqq$ and $qqqq\bar q$ 
components that arises from their unequal size 
parameters $\omega_3$
($qqq$) configuration and $\omega_5$ ($qqqq\bar q$)
has been taken into account by the following
factors:
\beq
\langle \phi_{000}({\vec\xi}_i)\vert 
\varphi_{000}({\vec\xi}_i)\rangle
=({\omega_3\omega_5\over\omega^2})^{3/2},~~~~
\langle \phi_{000}({\vec\xi}_i)\vert 
\varphi_{200}({\vec\xi}_i)\rangle
=\sqrt{{3\over 2}}({\omega_3\omega_5\over\omega^2})^{3/2}
[({\omega_5\over\omega})^2-1]\, .
\label{normfac}
\eeq 
Here the notation $\omega=\sqrt{(\omega_3^2+\omega_5^2)/2}$ 
has been employed. 

From (\ref{sfan}) and (\ref{orban}) 
the helicity amplitude for the direct 
$qqqq\bar q\rightarrow qqq\gamma$ transition is
obtained as:
\beq
A_{{1\over 2}}^{(53)}=-{2\sqrt{3}\over 9}{e\over \sqrt{K_\gamma}}~
C_{O}^{(5q\rightarrow 3q)}(k_\gamma)\, .
\label{ampand}
\eeq
The magnitude of this helicity amplitude contribution
depends sensitively on the size parameters $\omega_3$
and $\omega_5$. With the parameter values above
the numerical value $A_{{1\over 2}}^{(a)}=-0.035    
/\sqrt{\mathrm{GeV}}$ for $Q^2=0$. This value
is somewhat smaller than that given by the
conventional $qqq$ model.

Because of the same spin-flavor structure of 
the wave functions of the proton and $N(1440)$ in their 
3-quark and 5-quark configurations,
the $qqq\rightarrow qqqq\bar q \gamma$ transition described by 
Fig. \ref{fig1}(b) also contributes to 
the decays $N(1440)^+\rightarrow p$, which is different from the 
the case of $\Delta(1232)$ decay in refs. \cite{delpion,delgamma},
where the $qqq\rightarrow qqqq\bar q \gamma$ transition
does not contribute to the resultant decay amplitudes. 
In the case of the $qqq\rightarrow qqqq\bar q \gamma$ 
transition the 
spin-flavor-color matrix element 
is the same as that for the $qqqq\bar q\rightarrow qqq \gamma$ 
transition (\ref{sfan}). The orbital matrix element is however
different and takes the form:
\beqa
C_{O}^{(3q\rightarrow 5q)}(k_\gamma)=\langle~ 
\psi_C(\{{\vec\xi_i}\})~\vert~
\delta({\vec r}_4 - {\vec r}_5)
e^{-i{\vec k}_\gamma\cdot {{\vec r}_4 + {\vec r}_5\over 2}}~\vert~\nonumber\\
{1\over\sqrt{2}}(\phi_{000}({\vec\xi}_1)\phi_{200}({\vec\xi}_2)+
\phi_{200}({\vec\xi}_1)\phi_{000}({\vec\xi}_2))[111]_C~\rangle,
\label{oan35}
\eeqa   
where $[111]_C$ denotes the color wave function of $N(1440)$ in the 
$qqq$ configuration. As only the term multiplied by 
the color wave function $C_3$ in $\psi_C(\{{\vec\xi_i}\})$ 
leads to non-zero overlap in Eq. (\ref{oan35}), one obtains
the result:
\beq
C_{O}^{(3q\rightarrow 5q)}(k_\gamma)={\sqrt{6}\over 4}
({\omega_3\omega_5\over\omega^2})^3
[({\omega_3\over\omega})^2-1]
{k_\gamma\over\omega_5}~e^{-{3\over 20}{k_\gamma^2\over\omega_5^2}}\,.
\label{orban35}
\eeq 
The helicity amplitude for the direct 
$qqq\rightarrow qqqq\bar q\gamma$ transition is then
obtained as:
\beq
A_{{1\over 2}}^{(35)}=-{2\sqrt{3}\over 9}{e\over \sqrt{K_\gamma}}~
C_{O}^{(3q\rightarrow 5q)}(k_\gamma)\, .
\label{ampand35}
\eeq
With the parameter values above
the numerical value $A_{{1\over 2}}^{(a)}=0.0390    
/\sqrt{\mathrm{GeV}}$ for $Q^2=0$. This is large and  
has the opposite sign to that of $qqqq\bar q\rightarrow qqq\gamma$ 
transition, which further reduces the corresponding value in the 
$qqq$ model.

Combination of the helicity amplitudes in Eqs. (\ref{ampand})
and (\ref{ampand35}) with that of the direct transitions between 
the 3-quark and 5-quark components in $N(1440)$ and nucleon
leads to the net helicity 
amplitude: 
\beqa
A_{\frac{1}{2}}=A_{p3}A_{r3}A_{\frac{1}{2}}^{(3q)}
+A_{p5}A_{r5}A_{\frac{1}{2}}^{(5q)}+A_{p3}A_{r5}A_{\frac{1}{2}}^{(53)}
+A_{p5}A_{r3}A_{\frac{1}{2}}^{(35)}.
\eeqa
Assuming again a 10\% and 30\% proportion of five quark component
in proton and $N(1440)$ and the phase $+1$ for both
five quark components leads at the photon point, $Q^2=0$,
to the result $A_{\frac{1}{2}}=-0.022/\sqrt{\mathrm{GeV}}$. This value
is smaller than that derived in the $qqq$ model by a factor 0.6.
Similar numerical results holds even when a proportion of 5-quark 
component in $N(1440)$ larger than 30\% is assumed.

\subsubsection{$q\bar q\rightarrow \gamma$ transitions 
trigqered by quark-quark interactions}

\begin{figure}[t]
\vspace{20pt} 
\begin{center}
\mbox{\epsfig{file=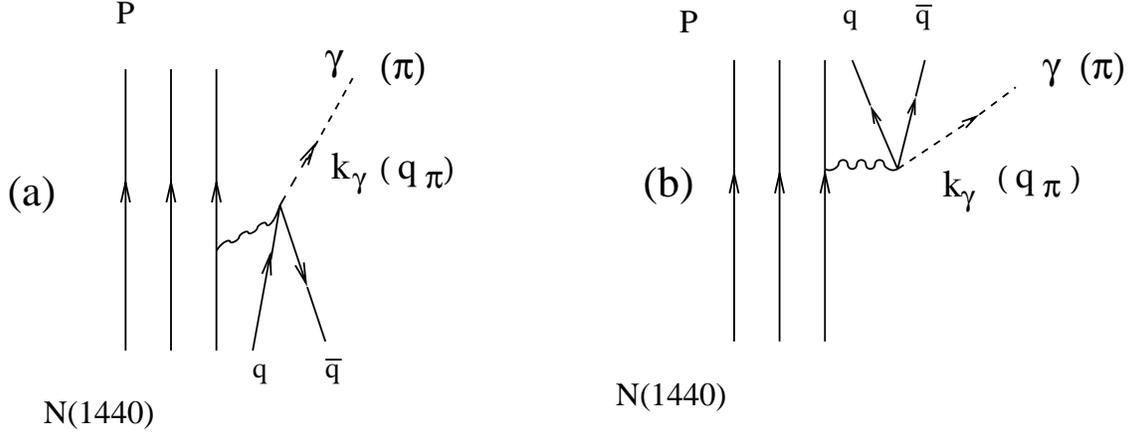, width=150mm}} 
\caption{Confinement induced
 $q\bar q\rightarrow \gamma (\pi)$ annihilation
process.}
\label{fig2}
\end{center}
\vspace{10pt} 
\end{figure}

Quark-antiquark annihilation transitions may
also be triggered
by the interactions between the quarks in the baryons.
The most obvious such triggering interaction is the
confining interaction, which is illustrated
diagrammatically in Fig. \ref{fig2}.
The role of this mechanism on the calculated
strong and electromagnetic  
decays of the $\Delta(1232)$ has been considered 
recently in Refs.
\cite{delpion,delgamma}.
The role of confinement driven annihilation on the
decays of the $N(1440)^+$ is considered here.

To lowest order in the quark momenta the amplitude for
this confinement triggered annihilation mechanism
may, in the case of a linear confining interaction,
be derived by means of the following replacement
in the transition operator for direct annihilation (\ref{at}):
\begin{equation}
e_i{\sigma_i}_-\rightarrow e_i{\sigma_i}_- 
(1-\frac{c r_{ij}-b}{2m})\,.
\label{shift}
\end{equation}  
Here c is the string tension, $r_{ij}$ is the
distance between the two quarks that interact by the
confining interaction and b is a constant, which shifts 
the zero point of the confinement to negative
values (b $>$ 0). This replacement
applies for both scalar and vector
coupled confinement. 
If the confining interaction is assumed to have the
color coupling $\vec\lambda_i^C\cdot\vec\lambda_j^C$,
the string tension $c$ should be the same for all the
$qq$ and $q\bar q$ pairs in the $qqqq\bar q$ system,
and half as strong as the corresponding
value for quark pairs in 
three-quark systems. Here we use the
value $c$ = 280 MeV/fm.

To derive the transition amplitude of the confinement
triggered annihilation mechanism one requires the 
matrix element of the transition operator
in orbital space, which, in the case of the 
$qqqq\bar q\rightarrow qqq\gamma$ described by 
Fig. \ref{fig2}(a), is defined as: 
\beq
X_{O}^{(5q\rightarrow 3q)}(k_\gamma)=\langle~ 
\phi_{000}({\vec\xi}_1)\phi_{000}({\vec\xi}_2)[111]_C~\vert~
\delta({\vec r}_4 - {\vec r}_5)~
e^{-i{\vec k}_\gamma\cdot {{\vec r}_4 + {\vec r}_5\over 2}}
~\bigg(\frac{c r_{34}-b}{2m}~\bigg)\vert~
\Psi_C(\{{\vec\xi_i}\})~\rangle.
\label{oanc}
\eeq 
Explicit evaluation leads to the result:
\beqa
X_{O}^{(5q\rightarrow 3q)}(k_\gamma)&=&
({\omega_3\omega_5\over\omega^2})^3~{64\sqrt{15}\over 375\pi}
{c\over m \omega}{k_\gamma\over\omega_5}
\{{\sqrt{6}\over 2}[({\omega_5\over\omega})^2-1]~K_0(k_\gamma)+
{\sqrt{6}\over 3}({\omega_5\over\omega})^2 K_1(k_\gamma)\nonumber\\
&+&{\sqrt{2}\over 2}K_2(k_\gamma)\}.
\label{orbanc}
\eeqa
Here the functions $K_i(q)$, i=0, 1, 2 have been defined as:
\beqa
&&K_0(q)=\omega_5^5\int_0^\infty d\xi\,\xi^4\,{j_1(\beta q\xi)
\over \beta q\xi}\,e^{-\alpha^2\xi^2}\, k_0(\omega\xi)\,,\nonumber\\
&&K_1(q)=\omega_5^5\int_0^\infty d\xi\,\xi^4\,{j_1(\beta q\xi)
\over \beta q\xi}\,e^{-\alpha^2\xi^2}\, k_1(\omega\xi)\, ,\nonumber\\
&&K_2(q)=\omega_5^7\int_0^\infty d\xi\,\xi^4\,
(X_1\xi^2-X_2{3\over{2\omega_5^2}})
{j_1(\beta q\xi)
\over \beta q\xi}\,e^{-\alpha^2\xi^2}\, k_0(\omega\xi)\, .
\label{K(xi)}
\eeqa
The constants
$\alpha$ and $\beta$ in this expression are defined
as $\alpha=2\omega_5/\sqrt{5}$ and $\beta=2\sqrt{3}/5$, 
respectively.
The functions $k_i(y)$, i=0, 1, is defined as:
\beqa
&& k_0(y)=\int_0^\infty dx x^2 e^{-x^2}\int_{-1}^1 dz
\{\sqrt{x^2
-2\sqrt{2} xzy+2y^2}-{\sqrt{6}\over 2}{b\omega\over c}\}\,  , \nonumber\\
&&k_1(y)=\int_0^\infty dx x^2 e^{-x^2}
(x^2-{3\over 2}{\omega^2\over \omega_5^2})
\int_{-1}^1 dz \{\sqrt{x^2
-2\sqrt{2} xzy+2y^2}-{\sqrt{6}\over 2}{b\omega\over c}\}\, .
\label{k(y)}
\eeqa

The space-flavor-color matrix element of the confinement driven
annihilation process is
the same 
as that of the direct annihilation given by (\ref{sfan}). 
Combining with the orbital matrix element in (\ref{orbanc}), one 
obtains the following contribution to the  
helicity amplitudes for $N(1440)\rightarrow N\gamma$ decay:
\beq
A_{{1\over 2}}^{(53c)}={2\sqrt{3}\over 3}{e\over \sqrt{K_\gamma}}~
X_{O}^{(5q\rightarrow 3q)}(k_\gamma).
\label{ampanc}
\eeq
Here an overall factor 3 has been inserted for
the 3 similar interacting processes
of annihilating the antiquark. At $Q^2=0$, the numerical value
of this contribution is large and positive in the
case that $b=0$: 
$A_{{1\over 2}}^{(53c)}=0.111/\sqrt{\mathrm{GeV}}$.

The orbital matrix element for the $qqq\rightarrow qqqq\bar q\gamma$
transition described by Fig. \ref{fig2}(b) is defined as 
\beqa
X_{O}^{(3q\rightarrow 5q)}(k_\gamma)=\langle~ 
\psi_C(\{{\vec\xi_i}\})~\vert~
\delta({\vec r}_4 - {\vec r}_5)
e^{-i{\vec k}_\gamma\cdot {{\vec r}_4 + {\vec r}_5\over 2}}~
\bigg(\frac{c r_{34}-b}{2m}~\bigg)~\vert \nonumber\\
{1\over\sqrt{2}}(\phi_{000}({\vec\xi}_1)\phi_{200}({\vec\xi}_2)+
\phi_{200}({\vec\xi}_1)\phi_{000}({\vec\xi}_2))[111]_C~\rangle,
\label{oanc35}
\eeqa
which leads explicitly to 
\beqa
X_{O}^{(3q\rightarrow 5q)}(k_\gamma)&=&
({\omega_3\omega_5\over\omega^2})^3~{128\sqrt{5}\over 375\pi}
{c\over m \omega}{k_\gamma\over\omega_5}
\{{3\over 2}[({\omega_3\over\omega})^2-1]~K_0(k_\gamma)+
({\omega_3\over\omega})^2 K_1^\prime(k_\gamma)\},
\label{orbanc35}
\eeqa
where
\beqa
&&K_1^\prime(q)=\omega_5^5\int_0^\infty d\xi\,\xi^4\,{j_1(\beta q\xi)
\over \beta q\xi}\,e^{-\alpha^2\xi^2}\, k_1^\prime(\omega\xi)\, ,\nonumber\\
&&k_1^\prime(y)=\int_0^\infty dx x^2 e^{-x^2}
(x^2-{3\over 2}{\omega^2\over \omega_3^2})
\int_{-1}^1 dz \{\sqrt{x^2
-2\sqrt{2} xzy+2y^2}-{\sqrt{6}\over 2}{b\omega\over c}\}\,. 
\label{K1prime}
\eeqa  

Combining with the orbital matrix element, one 
obtains the following contribution from the 
$qqq\rightarrow qqqq\bar q\gamma$ transition to the  
helicity amplitudes for $N(1440)\rightarrow N\gamma$ decay:
\beq
A_{{1\over 2}}^{(35c)}={2\sqrt{3}\over 3}{e\over \sqrt{K_\gamma}}~
X_{O}^{(3q\rightarrow 5q)}(k_\gamma).
\label{ampanc35}
\eeq
At $Q^2=0$, the numerical value
of this contribution is small and negative in the
case that $b=0$: 
$A_{{1\over 2}}^{(35c)}=-0.092/\sqrt{\mathrm{GeV}}$.

The role of the $q\bar q\rightarrow \gamma$ on the 
$N(1440)^+\rightarrow p\gamma$ decay naturally depends 
on the phase of the wave functions
of the $qqqq\bar q$ components of the proton and the $N(1440)$.
Combination of all the contributions above to the
helicity amplitude with 10\% $qqqq\bar q$ component in 
the proton and 30\% in
$N(1440)$ as above, but assuming a phase $-1$ for the 
$qqqq\bar q$ component, or $A_{p5}=-0.32$ and $A_{r5}=-0.55$
leads,  
with $b=0$, to a calculated value for the helicity amplitude  
$A_{{1\over 2}}$ of $-0.056$. This corresponds to 
an enhancement 
of the calculated decay width by a factor 2.3 over
that, which is derived in the $qqq$ quark model. 

\begin{figure}[t]
\vspace{20pt} 
\begin{center}
\mbox{\epsfig{file=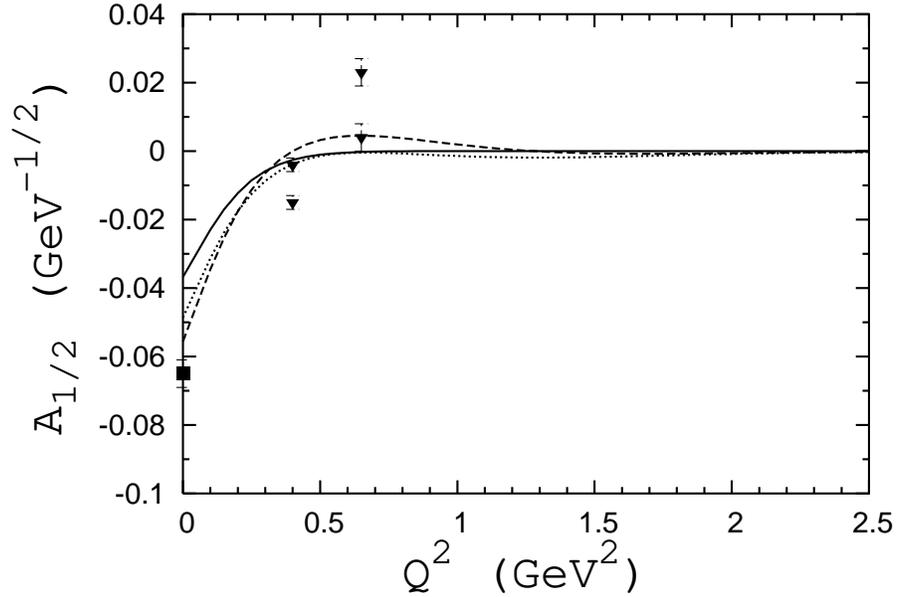, width=120mm}} 
\caption{The helicity amplitude for 
$N(1440)^+\rightarrow p\gamma$.
Solid line: The result for the conventional $qqq$ model;
dashed and dotted lines: the results when the amplitudes for the 
$qqqq\bar q\rightarrow qqqq \bar q$ transition and the direct 
and confinement triggered transitions through 
quark-antiquark annihilation are combined with the 
amplitude in the $qqq$ model with b=0 and b=200 MeV, 
respectively. Here the proportion of the 
5-quark components are assumed to be 10\% for the proton and 30\%
for the $N(1440)$, respectively. The data point at $Q^2 = 0$ is
from \cite{PDG} (square) and the other points are taken from the
phenomenological analysis in \cite{Azna} (triangles).}
\label{ampqq}
\end{center}
\vspace{10pt} 
\end{figure}

Hadron phenomenology suggests that the confining interaction
potential should be negative
at small distances \cite{timoc}. That is brought about by 
taking the value for $b$ to be positive. For instance, when 
$b=200$ MeV we get $A_{{1\over 2}}=-0.049$ at $Q^2=0$. 
The enhancement of the decay width 
over 
that in the $qqq$ model is virtually dominated by the confinement 
triggered annihilation transitions if $b<400$ MeV. For larger values 
of $b$ the contribution from the
confinement triggered annihilation amplitude
would serve to increase
the disagreement between the calculated and the
empirical values. To illustrate this the 
helicity amplitude is plotted in  Fig. \ref{ampqq}  
as a function of $Q^2$ both for the basic $qqq$
model and when the contributions 
from elastic $qqqq\bar q$ 
transition (\ref{hel5q}), the direct annihilation 
transitions and the confinement 
triggered annihilation transitions 
with $b=0$ and $b=200$ MeV are taken into account.

The qualitative features of the momentum dependence
of the calculated helicity amplitude $A_{1/2}$ in
Fig. \ref{ampqq} do not
depend very strongly on the presence of the $qqqq\bar q$
component, but the momentum behavior does follow the empirical
values somewhat better in the presence of those
components. The $qqqq\bar q$ component brings the value
at the photon point closer to the empirical value, and the
amplitude has a steeper rise at small values of
$Q^2$ and a change of sign of $A_{1\over 2}$, also in agreement 
with the recent empirical values \cite{Azna}. The magnitude of 
the calculated helicity amplitude in
all considered cases drops towards zero at about 1 GeV$^2$. 
The shape of the calculated helicity amplitude is rather 
insensitive to the precise value of $\omega_5$ (or to the
spatial extent of the $qqqq\bar q$ component). Increasing of
the numerical value of $\omega_5$ flattens the calculated curve. 
If the value of $\omega_5$ is
increased to 300 MeV the value of $A_{1/2}(0)$ is reduced
to -0.034 GeV$^{-1/2}$, but the shape of the calculated
function is affected only marginally.

\subsubsection{The role of other configurations in the $qqqq\bar q$
components}

Above the consideration of the $qqqq\bar q$ components was
restricted to the case, where the 4-quark
subsystem has flavor-spin symmetry
$[4]_{FS}[22]_F[22]_S$, and which is expected to have
the lowest energy of all $qqqq\bar q$ configurations. The
next to lowest energy $qqqq\bar q$ configuration in
both the nucleon and the $N(1440)$ is that
for which the flavor-spin symmetry is
$[4]_{FS}[31]_F[31]_S$ (\ref{q531}) \cite{helminen}.
As this configuration has the same color-space wave function
as the already considered configuration $[4]_{FS}[22]_F[22]_S$,
no significant change in the calculated results are 
to be expected by introduction of admixtures with the
flavor-spin symmetry $[4]_{FS}[31]_F[31]_S$.

Here only the $qqqq\bar q \rightarrow qqq$ and 
$qqq \rightarrow qqqq\bar q$ transitions for the $[4]_{FS}[31]_F[31]_S$ 
configuration are considered as 
the contributions from the transitions between five 
quark components in the $[4]_{FS}[22]_F[22]_S$ and $[4]_{FS}[31]_F[31]_S$ 
configurations are small.
The spin-flavor-color factor, that corresponds to
(\ref{sfan}) is obtained as,
\beq
C_{SFC}^{[31]}=-{10\sqrt{6}\over 27}\, ,
\label{sfan31}
\eeq 
both for the $qqqq\bar q \rightarrow qqq$ and 
$qqq \rightarrow qqqq\bar q$ transitions. This is larger than 
that for $[4]_{FS}[31]_F[31]_S$ (\ref{sfan}). The orbital matrix elements 
of annihilation transition amplitude
for the $[4]_{FS}[31]_F[31]_S$ 
configuration are the same as those of the $[4]_{FS}[22]_F[22]_S$   
configuration which are given by Eqs. (\ref{oan}), (\ref{oan35}), 
(\ref{oanc}) and (\ref{oanc35}).

\begin{figure}[t]
\vspace{20pt} 
\begin{center}
\mbox{\epsfig{file=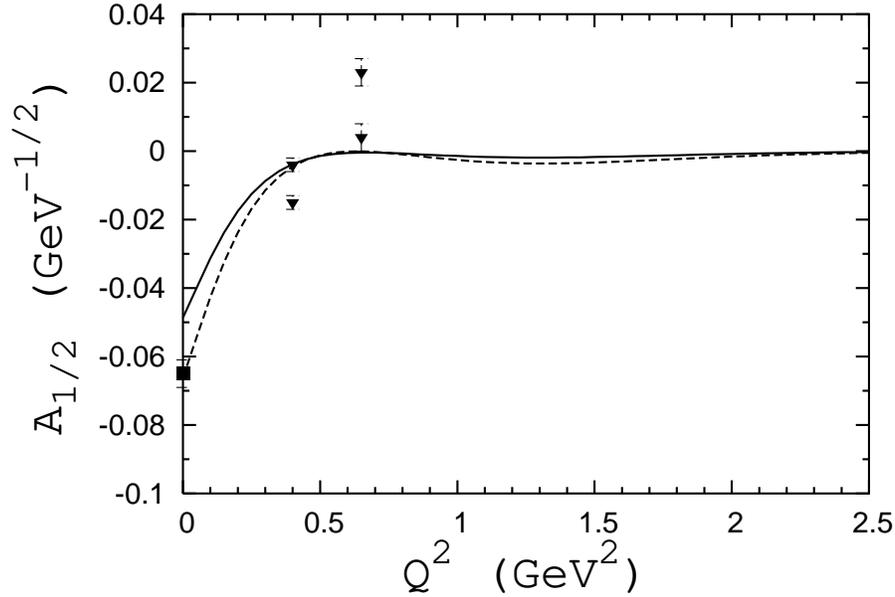, width=120mm}} 
\caption{The helicity amplitude of the decay 
$N(1440)^+\rightarrow p\gamma$ as a function of $Q^2$.
Solid line: only $[4]_{FS}[22]_F[22]_S$
configuration;
Dashed line: both $[4]_{FS}[22]_F[22]_S$
and $[4]_{FS}[31]_F[31]_S$ configurations are considered.
Here the probability of the 
$[4]_{FS}[22]_F[22]_S$ configuration in the $qqqq\bar q$
component is taken as be 80\%.
The data point at $Q^2 = 0$ is
from \cite{PDG} (square) and the other points are taken from the
phenomenological analysis in \cite{Azna} (triangles).}
\label{ampqq2con}
\end{center}
\vspace{10pt} 
\end{figure}

Here both the direct and confinement triggered
annihilation transitions $qqqq\bar q \rightarrow qqq$ and 
$qqq \rightarrow qqqq\bar q$ in the $[4]_{FS}[31]_F[31]_S$ 
configuration of proton and $N(1440)^+$ are considered as above.
Because the $[4]_{FS}[31]_F[31]_S$ configuration is expected to 
have a higher energy than the $[4]_{FS}[22]_F[22]_S$ configuration,
it should be expected to have a smaller probability than the 
latter configuration in the five quark component.  
With the same proportion of five quark component in proton and 
$N(1440)$ as before, but with 80\% of $[4]_{FS}[22]_F[22]_S$  
configuration and 20\% $[4]_{FS}[31]_F[31]_S$  configuration,
the final helicity amplitude is obtained as 
$A_{{1\over 2}}=-0.066/\sqrt{\mathrm{GeV}}$ with b=200 MeV at the 
real photon point. This value falls within the range of 
the empirical value \cite{PDG}. 
In Fig. \ref{ampqq2con} the calculated helicity 
amplitudes are shown with both configurations are included.
In comparison to 
the helicity amplitude with only the $[4]_{FS}[22]_F[22]_S$ 
configuration taken into account the new configuration does not 
lead to any qualitative 
change from the previous results, even though it does improve
the agreement quantitatively somewhat.

\section{The strong decay $N(1440)^+\rightarrow p\pi^0 $}

\subsection{The $qqq\rightarrow qqq\pi$ transition}

In the "chiral quark" model pions couple directly to
constituent quarks. In the non-relativistic
approximation the transition operator for 
$q\rightarrow q\pi^0$ 
approximation is then:
\begin{equation}
T_\pi = -i{g_A^q\over 2 f_\pi} \sum_i 
\tau_z ^i 
\sigma_z ^i\,q_{\pi z}\, .
\label{trans}
\end{equation}
Here the sum runs over the quarks, and $g_A^q$ is
the axial vector coupling constant of
the constituent quarks, $f_\pi$ the pion decay
constant and $q_{\pi z}$ the z-component of the 
pion momentum. 

With the conventional SU(6) $qqq$ wave functions (\ref{3qwave})
the transition operator for 
 $N(1440)^+ (s_z=1/2)
\rightarrow p(s_z=1/2)\pi^0$, the transition 
matrix element of the operator (\ref{trans}) in the valence
quark model is derived as:
\begin{equation}
T_\pi^{(3q)} = 
i{5\sqrt{3}\over 108}~{g_A^q\over f_\pi}~{{\vec q}_\pi\,^2\over
\omega_3^2}~e^{-\vec q_\pi\,^2/ 6\omega_3^2}~q_{\pi z}\, .
\label{3qm}
\end{equation}
Here the spatial
extent of the $qqq$ component of the baryon in the
harmonic oscillator model has been taken
into account. For the case of 
$N(1440)^+\rightarrow p\pi^0$ the pion momentum
is $q_\pi$ = 397 MeV.

The amplitude (\ref{3qm}) leads to values for the
decay width for $N(1440)\rightarrow N\pi$ that 
corresponds only to about $\sim 1/6$ of the
empirical value \cite{geb}. This deficiency is
a generic feature of the $qqq$ quark model
\cite{bruno, melde}.

\subsection{The $qqqq\bar q\rightarrow qqqq\bar q\pi$ transition}

The transition operator (\ref{trans}) leads to the
following contribution to the pion decay amplitude 
that connects the $qqqq\bar q$
components, in the $[4]_{FS}[22]_F[22]_S$ mixed symmetry
flavor-spin configuration, of the $N(1440)^+$ (\ref{5qr}) and 
the proton (\ref{5qn}): 
\begin{equation}
T_\pi^{(5q)} = 
-i{\sqrt{6}\over 180}~{g_A^q\over f_\pi}~{{\vec q}_\pi\,^2\over
\omega_5^2}~e^{-\vec q_\pi\,^2/ 5\omega_5^2}~q_{\pi z}\, .
\label{5qm}
\end{equation}
Note that, similarly to the case of the electromagnetic
decay of the $N(1440)^+$, the decay amplitude in (\ref{5qm}) only 
arises from the $\bar q\rightarrow \bar q\pi^0$
transition, while the quark 
transitions $q\rightarrow q\pi^0$ gives no contribution
in the configuration with $[22]$ spin symmetry.  

The transition amplitude for pion decay between 
the $qqqq\bar q$ components of $N(1440)$ and the
proton has the opposite sign 
to that given by the $qqq$ model. These $qqqq\bar q$ components
would therefore decrease the calculated final decay width
when the effect of these mechanisms are combined 
if the cross term matrix elements are not taken into account.
With a 10\% proportion 
of 5-quark component in proton and 30\% in $N(1440)$, 
these terms would imply a reduction
factor of 0.6 of the calculated decay width.

\subsection{The $q\bar q\rightarrow \pi$ transition}
\subsubsection{$q\bar q\rightarrow \pi$ transition without
 quark-quark interactions}
The simplest $qqqq\bar q \rightarrow qqq  \pi$
and $qqq\rightarrow qqqq\bar q \pi$
decay mechanisms that may contribute to the strong decay
of the $N(1440)$ are the $q\bar q\rightarrow \pi$
pair annihilation process which is shown in Fig. \ref{fig1}. 
The corresponding amplitude is
\begin{equation}
T_{\pi a} = -i{m g_A^q\over f_\pi}\tau_z,
\label{atp}
\end{equation} 
where $\tau_z$ represents the z-component of the
isospin of the annihilating quark.   

With the transition operator (\ref{atp}) the spin-flavor-color 
matrix element is derived from the 
wave functions of the proton in (\ref{3qwave}) and 
$N(1440)^+$ in (\ref{5qr}) as
\beq
C_{SFC}^{(5q\rightarrow 3q)}={\sqrt{3}\over 6}\, ,
\label{sfanp}
\eeq
in the case where the $qqqq\bar q$ component has flavor-
spin symmetry $[4]_{FS}[22]_F[22]_S$. 
Multiplication with the orbital matrix element 
(\ref{oan}) the final amplitude for the  
$N(1440)^+\rightarrow p\pi^0$ decay in the 
$qqqq\bar q\rightarrow qqq\pi$ mode is obtained as: 
\beq
T_{\pi}^{(53)}=-i{2\sqrt{3}\over 3}{m g_A^q\over f_\pi}~
C_{O}^{(5q\rightarrow 3q)}(q_\pi).
\label{ampandp}
\eeq
Here an overall factor 4 has been introduced to account for the 
numbers of quarks that may annihilate the antiquark.

Consider then the $qqq\rightarrow qqqq\bar q \pi$ 
transition described by Fig. \ref{fig1}(b).
The amplitude for the direct $qqq\rightarrow qqqq\bar q \pi$ 
transition is obtained as:
\beqa
T_{\pi}^{(35)}=-i{2\sqrt{3}\over 3}{m g_A^q\over f_\pi}~
C_{O}^{(3q\rightarrow 5q)}(q_\pi).
\label{ampandp35}
\eeqa
The matrix element in spin-flavor-color space is the
same in this case as in the case of the $qqqq\bar q\rightarrow qqq\pi$
transition in Eq. (\ref{sfanp}). The orbital matrix 
is given in Eq. (\ref{orban35}).  

As the amplitudes in (\ref{ampandp}) and  (\ref{ampandp35}) have 
the opposite sign to the one 
in the $qqq$ model, they would lead to an enhancement of the 
calculated 
decay width with the phase $-1$
in the wave functions of proton and $N(1440)$ 
as above. Again, with the assumption of 10\% and 30\% 
$qqqq\bar q$ components in proton and the Roper resonance 
respectively as above, these annihilation and creation
contributions lead to an enhancement factor of 
1.3 for the calculated decay width for $N(1440)\rightarrow p\pi^0$,
when the amplitude in (\ref{ampandp}) is combined with that of 
the transitions between the $qqq$ and $qqqq\bar q$
components. By itself this enhancement is not large enough
to compensate for the underprediction of the empirical
width in the quark model.

\subsubsection{$q\bar q\rightarrow \pi$ transition triggered by 
quark-quark interactions}
The strong decay of $N(1440)$
through quark-antiquark annihilation can also be triggered
by the interactions between the quarks in the baryons, as 
illustrated by Fig. \ref{fig2}. With the assumption of
a linear scalar coupled confining interaction  
between the quarks 
the amplitude for the confinement triggered annihilation mechanism
in the pion decay of $N(1440)$,
may to lowest order in the quark momenta be derived 
by making the following replacement
in the transition operator for direct annihilation (\ref{at}):
\begin{equation}
m\rightarrow m+{1\over 2}(c r_{ij}-b) \,.
\label{lconf}
\end{equation}  

With the spin-flavor-color matrix element (\ref{sfanp}) and 
the orbital matrix elements (\ref{oanc}) and (\ref{oanc35}) 
the
amplitudes for quark-antiquark 
annihilation triggered by the 
confining interaction is found to be
\beq
T_{\pi}^{(53c)}=-i2\sqrt{3}{m g_A^q\over f_\pi}~
X_{O}^{(5q\rightarrow 3q)}(q_\pi)\, ,
\label{ampancp}
\eeq 
for the $qqqq\bar q\rightarrow qqq\pi$ transition
and
\beq
T_{\pi}^{(35c)}=-i2\sqrt{3}{m g_A^q\over f_\pi}~
X_{O}^{(3q\rightarrow 5q)}(q_\pi)\,,
\label{ampancp35}
\eeq 
for the $qqq\rightarrow qqqq\bar q\pi$ transition.
Here an overall factor 12 has been introduced to account 
for the number of the similar processes of the 
confinement triggered annihilation of the quarks 
with the antiquark. 

The effect of the amplitude of the confinement 
triggered annihilation transition 
(\ref{ampancp}) on the final strong decay width 
of $N(1440)$ is very sensitive to the value of the
shift parameter  $b$ in the linear confining interaction.
With the value $b=$ 200 MeV, used above for the
helicity amplitude for electromagnetic decay, the 
the confinement 
triggered annihilation transition amplitude leads to an
enhancement of the calculated decay width  
of 6.1, when it is combined with the amplitudes of the transition between 
the $qqqq\bar q$ components and the direct annihilation transitions.
With $b=$ 0 the enhancement factor is 7.9.  
This enhancement is of the right order of magnitude to compensate
for the underprediction of the decay width in the conventional
$qqq$ quark model (\ref{3qm}).

\section{Discussion}

The result above shows that by explicit introduction of a
sizable, $\sim$ 30\%, $qqqq\bar q$ component in the low lying
broad $N(1440)$ resonance and a smaller such component
in the proton, it becomes possible to describe
the helicity amplitude $A_{1\over 2}$ for $N(1440)\rightarrow N\gamma$
decay and the empirical width for $N(1440)\rightarrow N\pi$
decay with the same set of quark model parameters. Without
such extension of the quark model it is difficult to
understand the observed structure and the large decay
width of the low lying
$N(1440)$ resonance. The fact that this resonance may
be described as a vibrational mode of the nucleon - i.e.
as a collective state - in the Skyrme model also suggests that
it should contain considerable multiquark components
\cite{bied}. This may also be indirectly inferred from the
fact that this resonance may be generated dynamically in
phenomenological hadronic coupled channel models \cite{krehl}.

With properly chosen size parameters of nucleon 
and $N(1440)$, the extension of the quark model to include $qqqq\bar q$
components considered here leads to a
transition form factor for the $N(1440)$ resonance, that
has the same strong momentum dependence that is
suggested by the phenomenological analysis of recent
data in ref.\cite{Azna}. This feature requires that the
wave function of the resonance be more extended spatially
than that of the nucleon.

Besides the $qqqq\bar q$ configuration 
with the isospin-spin symmetry $[4]_{FS}[22]_F[22]_S$, which
is expected to have the lowest energy because of the
spin dependence of the hyperfine interaction between the
quarks, other configurations such as $[4]_{FS}[31]_F[31]_S$
may also contribute to the final transition form factor 
quantitatively. The $[4]_{FS}[22]_F[22]_S$ symmetry is rather
restrictive, in that it only allows the antiquark in the
proton and the $N(1440)$ to be a $\bar d$ antiquark. The
$qqqq\bar q$ configuration with $[4]_{FS}[31]_F[31]_S$ 
flavor-spin symmetry 
is expected to have the next lowest
energy and would also allow
$\bar u$ antiquarks to be admixed into these baryons.
This symmetry configuration was also studied numerically
above, but was nevertheless found to bring no qualitative 
change to the numerical results.
The fact that empirically the $\bar d/ \bar u$
ratio is considerably larger than 1 in the proton
does nevertheless indicate that to a first
approximation, it may be sufficient to consider only the
configuration $[4]_{FS}[22]_F[22]_S$ \cite{towell}.


\begin{thebibliography}{99}

\bibitem{bruno} B. Juli\'a-D{\'i}az, D. O. Riska and
F. Coester, Phys. Rev. {\bf C70}, 045205 (2004)

\bibitem{melde} T. Melde, W. Plessas and R. F: Wagenbrunn, 
Phys. Rev. C72, 015207 (2005)

\bibitem{bied} L. C. Biedenharn, Y. Dothan and M. Tarlini,
Phys. Rev. {\bf D31}, 649 (1985)

\bibitem{helminen} C. Helminen and D. O. Riska, 
Nucl. Phys. {\bf A699}, 624 (2002) 

\bibitem{glozman} L. Ya. Glozman and D. O. Riska,
Phys. Rept. {\bf 268}, 263 (1996)

\bibitem{chen} J.-Q. Chen, Group Representation Theory for
Physicists, World Scientific, Singapore (1989)

\bibitem{PDG} S. Eidelman et al., Particle Data group,
Phys. Lett. {\bf 542}, 1 (2004)  

\bibitem{azna} I. Aznauryan et al., Phys. Rev {\bf C 72}, 045201 (2005)

\bibitem{delpion} Q. B. Li and D. O. Riska, Phys. Rev.
{\bf C 73}, 035201 (2006)

\bibitem{delgamma} Q. B. Li and D. O. Riska, Nucl. Phys.
{\bf A766}, 172 (2005)

\bibitem{timoc} K. O. Henriksson et al., Nucl. Phys. {\bf A674}, 141
(2000) 

\bibitem{Azna} I. Aznauryan et al., Phys, Rev. {\bf C71}, 
015201 (2005)

\bibitem{geb} D. O. Riska and G. E. Brown, Nucl. Phys.
{\bf A 679}, 577 (2001)

\bibitem{krehl} O. Krehl et al., Phys. Rev. {\bf C62}, 025207 (2000)

\bibitem{towell} R. S. Towell et al., Phys. Rev. {\bf D64},
052002 (2001)

\end{thebibliography}
\end{document}